\documentclass[12pt]{article}
\usepackage{amssymb, latexsym, amsmath, hyperref}
\usepackage{epsfig}
\usepackage{graphicx}
\usepackage{amsfonts}
\usepackage{mathrsfs}
\usepackage{multirow}
\usepackage{wrapfig}
\usepackage[dvips]{color}

\newcommand{\be}{\begin{equation}}
\newcommand{\ee}{\end{equation}}
\newcommand{\bea}{\begin{eqnarray}}
\newcommand{\eea}{\end{eqnarray}}

\newcommand{\ed}{\end{document}}

\newcommand{\bi}{\begin{itemize}}
\newcommand{\ei}{\end{itemize}}

\newcommand{\bce}{\begin{center}}
\newcommand{\ece}{\end{center}}

\begin{document}

\title{Euclidean Pseudoduality and Boundary Conditions in Sigma Models}

\author{Mustafa Sar{\i}saman\thanks{E-mail address:
msarisaman@ku.edu.tr, Phone: +90 212 338 1378, Fax: +90 212 338
1559}
\\
Department of Mathematics, Ko\c{c} University, \\ 34450 Sar{\i}yer,
Istanbul, Turkey}

\date{ }
\maketitle

\begin{abstract}
We discuss pseudoduality transformations in two dimensional
conformally invariant classical sigma models, and extend our
analysis to a given boundaries of world-sheet, which gives rise to
an appropriate framework for the discussion of the pseudoduality
between D-branes. We perform analysis using the Euclidean spacetime
and show that structures on the target space can be transformed into
pseudodual manifold identically. This map requires that torsions and
curvatures related to individual spaces are the same when
connections are riemannian. Boundary pseudoduality imposes locality
condition. \vspace{2mm}

\vspace{2mm}

\end{abstract}

\section{Introduction}\label{sec:int}

The term `duality' is widely used in physics literature to express
that two different systems turn out to be equivalent when there is a
duality transformation between these systems. In string theory
people use the term `target space duality' \cite{giveon, lizzi,
alvarez0, rahn} if there is a canonical transformation between
target spaces in which strings move. This transformation preserves
the hamiltonian.

In recent years a new type of duality transformation called
pseudoduality was suggested by Curtright and Zachos, Ivanov and
Alvarez \footnote{The term `pseudoduality' was first introduced by
Curtright and Zachos in \cite{curtright1}, used by Ivanov
\cite{ivanov}, and developed by Alvarez \cite{alvarez1,alvarez2}}
\cite{alvarez1, alvarez2, curtright1, ivanov}. This new topical
issue is quite interesting since it addresses duality transformation
on the world-sheet as distinct from the usual duality transformation
on the target space. The prominent feature of pseudoduality is to
preserve the stress-energy tensor, and therefore in principle not a
canonical transformation \cite{alvarez1, alvarez2}. This `on-shell'
duality transformation is carried out by mappings between the
solutions of the equations of motion.

In our prior research we analysed pseudoduality in symmetric space
sigma models \cite{msarisaman4} based on Lie group valued fields,
and extended it to supersymmetric case in \cite{msarisaman1,
msarisaman2, msarisaman3}. In these papers, there were some global
problems traced back to the signature of the worldsheet, especially
in supersymmetrized worldsheet. In order to designate and solve this
concern, we work with Wick rotated worldsheet in the present paper.

Recent studies \cite{alvarez1, msarisaman1, msarisaman2,
msarisaman3} about pseudoduality in sigma models revealed that
constructing pseudoduality in the worldsheet with lorentzian
signature is not a pleasant approach since the negative sign in the
pseudoduality expressions arising from lorentzian point of view
leads to the vanishing torsion in both manifolds and more
importantly when supersymmetry is imposed, to non invertible mapping
which maps all the points on one manifold to only one point on the
pseudodual manifold. This is the point that has vanishing Riemann
connection. It is understood \cite{msarisaman1} that it is better to
perform the pseudoduality between worldsheets which have general
(non-Lorentzian) signatures. To realize this goal, in this paper we
set up a simplified version, the Euclidean pseudoduality
transformations. Accordingly we extend our analysis to the given
boundaries of the world-sheet coordinates in the classical sigma
models.

It was observed \cite{msarisaman1} that expressing pseudoduality in
standard lightcone coordinates causes the geometry of target spaces
to be torsion free, and sigma model is not globally defined on the
pseudodual manifold. Therefore it is not invertible. To fix this
problem we will introduce alternative pseudoduality expressions in
the Euclidean worldsheet which is parameterized by $\tau$ and
$\sigma$. We point out that alternative pseudodualities adjust the
curvatures of target spaces by means of a modified connection so
that target spaces are globally well-defined and constructed as
diffeomorphic and dual symmetric spaces respectively with respect to
the modified connections in these cases, which are not just a
characteristic of sigma models. Pseudoduality also imposes that
torsions of the target spaces on which sigma models are based
produce infinitely many conditions related to their covariant
derivatives in the first case, and vanish in case of symmetric
spaces which coincides with the results obtained in literature.

\section{The Framework}

The sigma model with target space $M$, metric $g$ and antisymmetric
2-form $b$ is denoted by $(M, g, b)$ and has the action in the
Euclidean worldsheet $\Sigma$ \cite{hull1}

\begin{align}
S &= \frac{1}{2} \int_{\Sigma} d^{2} \sigma \sqrt{h}[h^{\mu \nu}
\partial_{\mu} x^{i} \partial_{\nu} x^{j}g_{i j} (x) + i \epsilon^{\mu \nu} \partial_{\mu} x^{i} \partial_{\nu} x^{j}b_{i j} (x)] \notag\\ & = \int_{\Sigma} d^{2} \sigma [(\frac{1}{2} g_{ij} \frac{\partial x^{i}} {\partial \tau}
\frac{\partial x^{j}} {\partial \tau} + \frac{1}{2} g_{ij}
\frac{\partial x^{i}} {\partial \sigma} \frac{\partial x^{i}}
{\partial \sigma}) + i b_{ij} \frac{\partial x^{i}} {\partial \tau}
\frac{\partial x^{j}} {\partial \sigma})] \label{equation1}
\end{align}
where $x : \Sigma \longrightarrow M$ is specified locally by the
functions $x^{i}(\sigma)$ giving the dependence of coordinates
$x^{i}$ of $M$ on the coordinates $\sigma^{\mu}$ of $\Sigma$. The
worldsheet $\Sigma$ is endowed with the Euclidean metric $h_{\mu
\nu}$ with $h = \mid \textrm{det}(h_{\mu \nu}) \mid$. The globally
defined closed 3-form $H$ is locally given by $H = db$. Notice that
Euclidean version of the action is obtained by Wick rotation of the
Lorentzian case (see appendix) in the case in which $h$ is flat.
Since $g_{i j} (x)$ and $b_{i j} (x)$ have real components, the term
involving $b$ is pure imaginary so that action is complex.

We will assume that sigma model is defined on a region $U$ of
$\Sigma$ with boundary $\partial U$. Equations of motion following
from this action in the bulk space will be

\begin{equation}
x_{\tau \tau}^{i} + x_{\sigma \sigma}^{i} = - \Gamma_{jk}^{i}
(x_{\tau}^{j} x_{\tau}^{k} + x_{\sigma}^{j} x_{\sigma}^{k}) + i
H_{jk}^{i} x_{\tau}^{j} x_{\sigma}^{k} \label{equation2}
\end{equation}
with the corresponding Dirichlet and Neumann boundary conditions respectively

\begin{align}
\delta x^{i} = 0 \label{equation3}\\
x_{\sigma}^{i} - i b_{j}^{i} x_{\tau}^{j} = 0 \label{equation4}
\end{align}
where we defined $x_{\tau} \equiv \frac{\partial x}{\partial \tau}$
and $x_{\sigma} \equiv \frac{\partial x}{\partial \sigma}$. We would
like to relate the sigma model $(M, g, b)$ to a different one
$(\tilde{M}, \tilde{g}, \tilde{b})$ by means of these equations. The
pseudodual model will be represented by $(\tilde{M}, \tilde{g},
\tilde{b})$ and similar expressions may be written on $(\tilde{M},
\tilde{g}, \tilde{b})$ using tilde. As it is well known
pseudoduality equations are best formulated on the orthonormal
coframe bundle \footnote{Orthonormal coframe bundle is defined on
$SO(M) = M \times SO(n)$.}.

We choose an orthonormal frame $\{\theta^{i}\}$ with the riemannian connection $\omega_{j}^{i}$ defined on the worldsheet as

\begin{equation}
\theta^{i} = x_{a}^{i} d\pi^{a} = x_{\tau}^{i} d \tau +
x_{\sigma}^{i} d \sigma \label{equation5}
\end{equation}
where the worldsheet coordinates are given by $\pi = (\tau,
\sigma)$. The indices in the middle of alphabet denote coordinates
on the target manifold while indices $a, b, c...$ represent
coordinates of the worldsheet. In what follows we will construct two
different pseudoduality equations. The first one yields the
coordinate diffeomorphisms and the second case restricts manifolds
to symmetric spaces.

\subsection{Case I: Pseudoduality to Coordinate Diffeomorphisms}

In this case we assume that pseudoduality equations are defined on
wick rotated world-sheet using the pullback bundle of target space
as $\tilde{\theta} = T \theta$. These are explicitly written as
\begin{align}
\tilde{x}_{\tau}^{i} = T_{j}^{i} x_{\tau}^{j} \label{equation6}\\
\tilde{x}_{\sigma}^{i} = T_{j}^{i} x_{\sigma}^{j} \label{equation7}
\end{align}
in order to better understand these equations we will inquire the integrability conditions of these equations as in \cite{alvarez1, msarisaman1}. We define the covariant derivative of $x_{a}^{i}$
\begin{equation}
dx_{a}^{i} + \omega_{j}^{i} x_{a}^{j} = x_{ab}^{i} d\pi^{b} \label{equation8}
\end{equation}
Cartan structural equations are given by
\begin{align}
d\theta^{i} + \omega_{j}^{i} \wedge \theta^{j} \label{equation9}&= 0 \\
d\omega_{j}^{i} + \omega_{k}^{i} \wedge \omega_{j}^{k} &=
\Omega_{j}^{i} \label{equation10}
\end{align}
where $\Omega_{j}^{i} = \frac{1}{2} R_{jkl}^{i} \theta^{k} \wedge
\theta^{l}$ is the curvature two-form. We take the exterior
derivative of (\ref{equation6}) and (\ref{equation7}) and use the
covariant derivative (\ref{equation8}) to obtain the following
    \begin{align}
    \tilde{x}_{\tau b}^{i} d\pi^{b} &= [dT_{k}^{i} - T_{j}^{i} \omega_{k}^{j} + \tilde{\omega}_{j}^{i} T_{k}^{j}]x_{\tau}^{k} + T_{j}^{i} x_{\tau b}^{j} d
    \pi^{b}\label{equation11}\\
    \tilde{x}_{\sigma b}^{i} d\pi^{b} &= [dT_{k}^{i} - T_{j}^{i} \omega_{k}^{j} + \tilde{\omega}_{j}^{i} T_{k}^{j}]x_{\sigma}^{k} + T_{j}^{i} x_{\sigma b}^{j} d
    \pi^{b} \label{equation12}
    \end{align}
These two equations are intriguing and lead to take advantage of
equations of motion and the desired integrability conditions for the
pseudoduality equations. As opposed to the method followed in
\cite{msarisaman1, msarisaman2} we first and foremost wedge the
first equation (\ref{equation11}) by $d\tau$ and the second
(\ref{equation12}) by $d\sigma$, and use $x_{\tau \sigma} =
x_{\sigma \tau}$ (similarly $\tilde{x}_{\tau \sigma} =
\tilde{x}_{\sigma \tau}$) to get
     \begin{equation}
     [dT_{k}^{i} - T_{j}^{i} \omega_{k}^{j} + \tilde{\omega}_{j}^{i}
     T_{k}^{j}]x_{\tau}^{k} \wedge d\tau + [dT_{k}^{i} - T_{j}^{i} \omega_{k}^{j} + \tilde{\omega}_{j}^{i}
     T_{k}^{j}]x_{\sigma}^{k} \wedge d\sigma = 0 \label{equation13}
     \end{equation}
For the sake of clarity we split the core part of this equation as
$[dT_{k}^{i} - T_{j}^{i} \omega_{k}^{j} + \tilde{\omega}_{j}^{i}
     T_{k}^{j}] = \mathcal{A}_{k\tau}^{i} d\tau + \mathcal{B}_{k\sigma}^{i}
     d\sigma$. Therefore, with the use of (\ref{equation5}), (\ref{equation13}) can be rewritten as
     \begin{equation}
     \mathcal{C}_{k a}^{i} \theta^{k} \wedge d\pi^{a} := \mathcal{A}_{k\tau}^{i} \theta^{k} \wedge d\tau +
     \mathcal{B}_{k\sigma}^{i} \theta^{k} \wedge d\sigma = 0 \notag
     \end{equation}
We consider first the ``weak'' case where $\mathcal{C}_{k a}^{i} =
0$, which requires $\mathcal{A}_{k\tau}^{i} = 0$ and
$\mathcal{B}_{k\sigma}^{i} = 0$. Therefore we come up with the first
integrability condition
      \begin{equation}
      [dT_{k}^{i} - T_{j}^{i} \omega_{k}^{j} + \tilde{\omega}_{j}^{i}
     T_{k}^{j}] = 0 \label{equation14}
      \end{equation}
Notice that we have not still made use of the equations of motion
for sigma models. This leads to the conclusion that
(\ref{equation14}) is not special to just sigma models but a
property of pseudoduality itself. What characterizes pseudoduality
in sigma models is obtained by wedging (\ref{equation11}) by
$d\sigma$ and (\ref{equation12}) by $d\tau$, and subtracting the
resulting equations to get
       \begin{equation}
       (\tilde{x}_{\tau \tau}^{i} + \tilde{x}_{\sigma \sigma}^{i}) d\tau \wedge d\sigma
       = T_{j}^{i} (x_{\tau \tau}^{j} + x_{\sigma \sigma}^{j}) d\tau \wedge d\sigma
       \label{equation15}
       \end{equation}
$d\tau \wedge d\sigma$ is trivial and can be cancelled out.
Inserting the equations of motion (\ref{equation2}) one obtains
       \begin{equation}
       - \tilde{\Gamma}_{jk}^{i}
(\tilde{x}_{\tau}^{j} \tilde{x}_{\tau}^{k} + \tilde{x}_{\sigma}^{j}
\tilde{x}_{\sigma}^{k}) + i \tilde{H}_{jk}^{i} \tilde{x}_{\tau}^{j}
\tilde{x}_{\sigma}^{k} = - T_{j}^{i} \Gamma_{m n}^{j} (x_{\tau}^{m}
x_{\tau}^{n} + x_{\sigma}^{m} x_{\sigma}^{n}) + i T_{j}^{i} H_{m
n}^{j} x_{\tau}^{m} x_{\sigma}^{n} \label{equation16}
       \end{equation}
Notice that this equation consists of symmetric and antisymmetric
parts, which leads to the decomposition into two distinct equations.
Using pseudoduality equations (\ref{equation6}) and
(\ref{equation7}) in these resulting expressions yields the
remaining integrability conditions which are special to sigma models
       \begin{align}
       T_{j}^{i} \Gamma_{m
       n}^{j} &= \tilde{\Gamma}_{jk}^{i} T_{m}^{j} T_{n}^{k}  \label{equation17}\\
       T_{j}^{i} H_{m
n}^{j} &= \tilde{H}_{jk}^{i} T_{m}^{j} T_{n}^{k} \label{equation18}
       \end{align}
These equations can be investigated further by taking exterior
derivatives. Exterior derivative of (\ref{equation17}) together with
condition (\ref{equation14}) yields that
      \begin{equation}
      T_{j}^{i} \Omega_{n}^{j} = \tilde{\Omega}_{j}^{i} T_{n}^{j}
      \label{equation19}
      \end{equation}
where we defined $\omega_{j}^{i} := \Gamma_{k j}^{i} \theta^{k}$,
$\Gamma_{j}^{i} := \Gamma_{k j}^{i} \theta^{k}$ and used the
curvature two form $d\Omega_{j}^{'i} = d\Omega_{j}^{i} =
d\Gamma_{j}^{i} + \Gamma_{k}^{i} \wedge \Gamma_{j}^{k}$ (similarly
for $\tilde{\omega}_{j}^{i} :=\tilde{\Gamma}_{k j}^{i}
\tilde{\theta}^{k}$, $\tilde{\Gamma}_{j}^{i} := \tilde{\Gamma}_{k
j}^{i} \tilde{\theta}^{k}$ and $d\tilde{\Omega}_{j}^{i} =
d\tilde{\Omega}_{j}^{'i} = d\tilde{\Gamma}_{j}^{i} +
\tilde{\Gamma}_{k}^{i} \wedge \tilde{\Gamma}_{j}^{k}$ ). This
requires that curvatures of pseudodual manifolds are related to each
other by
      \begin{equation}
      T_{j}^{i} R_{k\ell m}^{j} = \tilde{R}_{j\ell p}^{i} T_{k}^{j}
      T_{m}^{\ell} T_{n}^{p} \label{equation20}
      \end{equation}
One may continue taking additional exterior derivatives to
understand the integrability conditions. But notice that
(\ref{equation17}) causes (\ref{equation14}) to reduce to the form
$dT = 0$, which yields that $T$ is constant. Hence it is understood
that $T$ is just a constant change of bulk coordinates, $x^{i} =
T_{j}^{i} x^{j}$, which gives an obvious interpretation of the
equations (\ref{equation17}), (\ref{equation18}) and
(\ref{equation20}). A constant change of coordinates is an obvious
pseudoduality, since the sigma-model is invariant under bulk
diffeomorphisms. Therefore, torsions and curvatures together with
their covariant derivatives are equivalent to each other as
expected, i.e.
     \begin{align}
     R_{jk\ell}^{i} &= \tilde{R}_{jk\ell}^{i},~~~~~~DR_{jk\ell}^{i} = \tilde{D}\tilde{R}_{jk\ell}^{i},~~~~~~D\ldots DR_{jk\ell}^{i} = \tilde{D}\ldots
     \tilde{D}\tilde{R}_{jk\ell}^{i}, \notag\\
     H_{jk}^{i} &= \tilde{H}_{jk}^{i},~~~~~~~DH_{jk}^{i} = \tilde{D}\tilde{H}_{jk}^{i},~~~~~~~~D\ldots DH_{jk}^{i} = \tilde{D}\ldots
     \tilde{D}\tilde{H}_{jk}^{i}, \label{eq21}
     \end{align}
where $D = \tilde{D}$ is the covariant derivative with respect to
the associated riemannian connection.

If one uses the general case that the components of $\mathcal{C}_{k
a}^{i}$ are related to each other by the relation
(\ref{equation13}), and follow the same steps as above one obtains
the following conditions
\begin{align}
&T_{j}^{i} H_{mn}^{j} = \tilde{H}_{jk}^{i} T_{m}^{j} T_{n}^{k} \label{equation21}\\
&d T_{j}^{i} + \tilde{\omega}_{k}^{i} T_{j}^{k} - T_{k}^{i}
\omega_{j}^{k} +  T_{k}^{i} \Gamma_{lj}^{k} \theta^{l} -
\tilde{\Gamma}_{kl}^{i} T_{m}^{k} T_{j}^{l} \theta^{m} = 0
\label{equation22}
\end{align}
The first result is the same as (\ref{equation18}) and reveals the
relation between torsions. The main difference with the ``weak''
case comes with the second relation and in order to better
understand it we define a new (modified) connection $\xi_{j}^{i}:=
\omega_{j}^{i} - \Gamma_{kj}^{i} \theta^{k}$ (and
$\tilde{\xi_{j}^{i}}:= \tilde{\omega}_{j}^{i} -
\tilde{\Gamma}_{kj}^{i} \tilde{\theta}^{k}$), which leads
(\ref{equation22}) to
\begin{equation}
d T_{j}^{i} + \tilde{\xi}_{k}^{i} T_{j}^{k} - T_{k}^{i} \xi_{j}^{k}
= 0 \label{equa23}
\end{equation}
Hence it is manifest that the ``weak'' case corresponds to $\xi=
\tilde{\xi} = 0$. Since the characteristics of the pseudoduality is
encoded in the transformation map $T$, it is required to further
seek out the integrability of (\ref{equa23}), which produces that
      \begin{equation}
      T_{j}^{i} (\Omega_{\xi})_{k}^{j} = (\tilde{\Omega}_{\xi})_{j}^{i} T_{k}^{j}
      \label{equa24}
      \end{equation}
where we defined a new (modified) curvature two form\footnote{Cartan
structural equation for this modified connection in terms of
modified curvature two form is $d\xi_{j}^{i} + \xi_{k}^{i} \wedge
\xi_{j}^{k} = (\Omega_{\xi})_{j}^{i}$. Substituting $\xi_{j}^{i}=
\omega_{j}^{i} - \Gamma_{kj}^{i} \theta^{k}$ in this equation one
obtaines the corresponding Cartan structural equations for
$\omega_{j}^{i}$ and $\Gamma_{j}^{i}$.} $(\Omega_{\xi})_{j}^{i} :=
\Omega_{j}^{i} - \Omega_{j}^{'i} := \frac{1}{2}(R_{\xi})_{j k
\ell}^{i} \theta^{k} \wedge \theta^{\ell}$ (and the same relation
with tilde on pseudodual manifold). Therefore, the relation between
curvatures is similar to (\ref{equation20}) and given by
      \begin{equation}
      T_{j}^{i} (R_{\xi})_{k m n}^{j} = (\tilde{R}_{\xi})_{j\ell p}^{i} T_{k}^{j}
      T_{m}^{\ell} T_{n}^{p} \label{equa25}
      \end{equation}
One can work out further integrability of (\ref{equa25}) using
(\ref{equa23}) to obtain that
      \begin{equation}
      T_{j}^{i} (D_{\xi}R_{\xi})_{k\ell m}^{j} = (\tilde{D}_{\tilde{\xi}}\tilde{R}_{\xi})_{j\ell p}^{i} T_{k}^{j}
      T_{m}^{\ell} T_{n}^{p}, ~~~~~~~~ T_{j}^{i} (D_{\xi}\ldots D_{\xi}R_{\xi})_{k\ell m}^{j} = (\tilde{D}_{\tilde{\xi}}\ldots \tilde{D}_{\tilde{\xi}}\tilde{R}_{\xi})_{j\ell p}^{i} T_{k}^{j}
      T_{m}^{\ell} T_{n}^{p}\label{eq26}
      \end{equation}
where $D_{\xi}$ ($\tilde{D}_{\tilde{\xi}}$) is the covariant
derivatives with respect to the modified connection $\xi$ (
$\tilde{\xi}$), and defined by
      \begin{equation}
      (D_{\xi} R_{\xi})_{jk\ell}^{i} := d (R_{\xi})_{jk\ell}^{i} +
      (R_{\xi})_{jk\ell}^{q} \xi_{q}^{i} - (R_{\xi})_{qk\ell}^{i}
      \xi_{j}^{q}-(R_{\xi})_{jq\ell}^{i} \xi_{k}^{q}-(R_{\xi})_{jkq}^{i}
      \xi_{\ell}^{q} \notag
      \end{equation}
Therefore, we obtain the conclusion that pseudoduality in general
sense requires connections $\xi$ and $\tilde{\xi}$ defined
respectively on manifolds $M$ and $\tilde{M}$ to be related to each
other by the psedudoduality relation (\ref{equa23}). Unlike the
cases we discussed earlier, curvatures are not constant and the
same, but newly defined (modified) curvatures related to modified
connections are preserved under the map $T$. Compared to results
found in \cite{alvarez1, alvarez2, msarisaman1, msarisaman2,
msarisaman3, msarisaman4} this does not amount to symmetric spaces
with respect to modified spaces with connections $\xi$ and
$\tilde{\xi}$. In the special case that modified connection vanishes
(``weak'' case above), usual curvature relations are obtained.

To interpret torsions similarly, one needs to take exterior
derivative of (\ref{equation21}), which leads to
      \begin{equation}
      T_{j}^{i} (D_{\xi}H_{mn}^{j}) = (\tilde{D}_{\tilde{\xi}}\tilde{H}_{jk}^{i}) T_{m}^{j}
      T_{n}^{k} \label{equa26}
      \end{equation}
where the covariant derivative $D_{\xi}$ of $H_{jk}^{i}$ with
respect to $\xi$ is defined by
      \begin{equation}
      D_{\xi}H_{jk}^{i} := d H_{jk}^{i} + H_{jk}^{q}\xi_{q}^{i} -
      H_{qk}^{i}\xi_{j}^{q} - H_{jq}^{i}\xi_{k}^{q} \notag
      \end{equation}
Taking further exterior derivatives by repeated use of
(\ref{equa23}) produces infinitely many integrability conditions in
terms of covariant derivatives with respect to $\xi$ and
$\tilde{\xi}$ for $H_{jk}^{i}$ and $\tilde{H_{jk}^{i}}$
      \begin{align}
      T_{j}^{i} (D_{\xi}D_{\xi}H_{mn}^{j}) &= (\tilde{D}_{\tilde{\xi}}\tilde{D}_{\tilde{\xi}}\tilde{H}_{jk}^{i}) T_{m}^{j}
      T_{n}^{k} \notag\\
      T_{j}^{i} (D_{\xi}D_{\xi}D_{\xi}H_{mn}^{j}) &= (\tilde{D}_{\tilde{\xi}}\tilde{D}_{\tilde{\xi}}\tilde{D}_{\tilde{\xi}}\tilde{H}_{jk}^{i}) T_{m}^{j}
      T_{n}^{k} \notag\\
       T_{j}^{i} (D_{\xi}\ldots D_{\xi}H_{mn}^{j}) &= (\tilde{D}_{\tilde{\xi}}\ldots \tilde{D}_{\tilde{\xi}}\tilde{H}_{jk}^{i}) T_{m}^{j}
      T_{n}^{k} \label{equa27}
      \end{align}
Therefore, torsions under pseudoduality are mapped by
(\ref{equation21}), (\ref{equa26}) and (\ref{equa27}).

In case of supersymmetric extension of the worldsheet, it is obvious
that one can find the same results if the methods and conventions in
\cite{msarisaman1} are followed. This will not be discussed here.

\subsection{Case II: Pseudoduality to Symmetric Spaces}

Based on above results, an alternative expression which will make
use of the equations of motion can be written for pseudoduality as
$\tilde{\theta} = _{\ast \Sigma} T \theta$, where $_{\ast \Sigma}$
is the Hodge duality operator, or explicitly
     \begin{align}
     \tilde{x}_{\tau}^{i} &= T_{j}^{i} x_{\sigma}^{j} \label{eq6}\\
     \tilde{x}_{\sigma}^{i} &= -T_{j}^{i} x_{\tau}^{j} \label{eq7}
     \end{align}
Notice that we put a negative sign in the second equation to satisfy
the equations of motion. After a little computation it is easy to
show that equations (\ref{equation11}) and (\ref{equation12}) turn
to
     \begin{align}
    \tilde{x}_{\tau b}^{i} d\pi^{b} &= [dT_{k}^{i} - T_{j}^{i} \omega_{k}^{j} + \tilde{\omega}_{j}^{i} T_{k}^{j}]x_{\sigma}^{k} + T_{j}^{i} x_{\sigma b}^{j} d
    \pi^{b}\label{eq11}\\
    \tilde{x}_{\sigma b}^{i} d\pi^{b} &= -[dT_{k}^{i} - T_{j}^{i} \omega_{k}^{j} + \tilde{\omega}_{j}^{i} T_{k}^{j}]x_{\tau}^{k} - T_{j}^{i} x_{\tau b}^{j} d
    \pi^{b} \label{eq12}
    \end{align}
Wedging first equation by $d\tau$ and second equation by $d\sigma$,
and adding together produces two results
    \begin{align}
    &H = 0 \label{eq43}\\
    &dT_{k}^{i} + \tilde{\omega}_{j}^{i} T_{k}^{j} - T_{j}^{i} \gamma_{k}^{j} = 0 \label{eq44}
    \end{align}
where we used the equations of motion for $(M, g, b)$ and defined a
modified connection $\gamma_{j}^{i} := \omega_{j}^{i} + \Gamma_{k
j}^{i}\theta^{k}$ on manifold $M$. Likewise, one wedges the first
equation by  $d\sigma$ and the second equation by $d\tau$, and
subtract from each other to produce the following results
    \begin{align}
    &\tilde{H} = 0 \label{eq45}\\
    &dT_{k}^{i} + \tilde{\gamma}_{j}^{i} T_{k}^{j} - T_{j}^{i} \omega_{k}^{j} = 0 \label{eq46}
    \end{align}
where we used the equations of motion for sigma model $(\tilde{M},
\tilde{g}, \tilde{b})$ and defined a new modified connection
$\tilde{\gamma}_{j}^{i} := \tilde{\omega}_{j}^{i} +
\tilde{\Gamma}_{k j}^{i}\theta^{k}$ on manifold $\tilde{M}$. Thus we
understand that this type of pseudoduality kills torsions $H$ and
$\tilde{H}$, and thus gives rise to torsionless manifolds. Actually
this explains why Ivanov used torsionless manifolds in his
construction \cite{ivanov}. To grasp the remaining equations we
subtract (\ref{eq44}) from (\ref{eq46}) to get
    \begin{equation}
    \tilde{\Gamma}_{j}^{i} T_{k}^{j} = - T_{j}^{i} \Gamma_{k}^{j}
    \label{eq47}
    \end{equation}
This impressive relation is similar to (\ref{equation17}) except for
the negative sign and actually determines the geometry of the
manifolds under pseudoduality. We take the exterior derivative and
selectively insert (\ref{eq44}) and (\ref{eq46}) to get
    \begin{equation}
    d\tilde{\Omega}_{j}^{i} T_{k}^{j} = -T_{j}^{i} d\Omega_{k}^{j}
    \notag
    \end{equation}
where we used $d\Omega_{j}^{i} = d \Gamma_{j}^{i} + \Gamma_{k}^{i}
\wedge \Gamma_{j}^{k}$ (and similarly for tilded expression). This
gives us a relation between curvatures similar to (\ref{equation20})
    \begin{equation}
    \tilde{R}_{jk\ell}^{i} T_{m}^{j} T_{n}^{k} T_{p}^{\ell}= -T_{j}^{i} R_{mnp}^{j}
    \label{eq48}
    \end{equation}
Intriguing point presents itself when we take one more exterior
derivative and again selectively insert (\ref{eq44}) and
(\ref{eq46}) to get $DR_{jk\ell}^{i} =
\tilde{D}\tilde{R}_{jk\ell}^{i} = 0$, where $D$ is the covariant
derivative with respect to the connection $\gamma_{j}^{i}$ and
$\tilde{D}$ is the covariant derivative with respect to
$\tilde{\gamma}_{j}^{i}$. Therefore, we obtain that manifolds are
symmetric spaces with opposite curvatures. We obtain similar result
as in \cite{msarisaman5} using a different version of pseudoduality
equations

In conclusion, we understand that case I pseudoduality arising from
the setup $\tilde{\theta} = T\theta$ yields a coordinate
diffeomorphism with modified connections, but case II pseudoduality
originating from $\tilde{\theta} = _{\ast \Sigma}T\theta$ leads
manifolds $M$ and $\tilde{M}$ torsionless and symmetric spaces with
opposite curvatures. It is obvious that symmetric space property is
a result of hodge duality operator. Unlike the results found in
\cite{alvarez1, alvarez2, msarisaman1, msarisaman2}, pseudoduality
imposes the restriction that manifolds are torsionless.

\section{Pseudoduality at Boundaries}

We extend our analysis around the boundaries of the region $\partial
U$ of the worldsheet $\Sigma$. Pseudoduality can be formulated at
boundaries by means of the Stokes' theorem, $\int_{\partial U}
\theta' = \int_{U} d \theta'$ where $d \theta' = \theta$ and
$\theta'$ is defined at the boundaries. As one might expect
pseudoduality equations are reduced to
\begin{equation}
\tilde{x}^{i} = T_{j}^{i} x^{j} \label{equation24}
\end{equation}
where we use Dirichlet, Neumann or mix mappings $T = \{T_{D}, T_{N},
T_{M}\}$ depending on the type of boundary conditions. If we only
have Dirichlet boundary condition (\ref{equation3}) then $T = T_{D}$
is a constant and pseudoduality equations are simply $\tilde{x}^{i}
= x^{i}$. If we only have the Neumann boundary condition
(\ref{equation4}), then taking the exterior derivative of
(\ref{equation24}) with $T = T_{N}$ gives
\begin{equation}
d\tilde{x}^{i} = (d T_{N})_{j}^{i} x^{j} + (T_{N})_{j}^{i} dx^{j}
\label{equation25}
\end{equation}
where $\theta' = x$ is taken at boundary $\partial U$. We define the
corresponding connection one form $\omega_{j}^{'i}$ and the
covariant derivative
\begin{equation}
d x^{i} + \omega_{j}^{'i} x^{j} = x_{a}^{i} d \pi^{a} \notag
\end{equation}
Substituting the covariant derivative in (\ref{equation25}) together
with (\ref{equation24}) we obtain
\begin{equation}
 \tilde{x}_{a}^{i} d \pi^{a} = [(d T_{N})_{k}^{i} + \tilde{\omega}_{j}^{'i} (T_{N})_{k}^{j} - (T_{N})_{j}^{i}\omega_{k}^{'j}] x^{k} +  (T_{N})_{j}^{i} x_{a}^{j} d \pi^{a}  \notag
\end{equation}
Subsequently we wedge this expression with $d\sigma$ and $d\tau$ to
obtain
\begin{align}
\tilde{x}_{\tau}^{i} d\tau \wedge d\sigma &= [(d T_{N})_{k}^{i} +
\tilde{\omega}_{j}^{'i} (T_{N})_{k}^{j} -
(T_{N})_{j}^{i}\omega_{k}^{'j}]
x^{k} \wedge d\sigma +  (T_{N})_{j}^{i} x_{\tau}^{j} d\tau \wedge d\sigma \notag \\
\tilde{x}_{\sigma}^{i} d\sigma \wedge d\tau &= [(d T_{N})_{k}^{i} +
\tilde{\omega}_{j}^{'i} (T_{N})_{k}^{j} -
(T_{N})_{j}^{i}\omega_{k}^{'j}] x^{k} \wedge d\tau + (T_{N})_{j}^{i}
x_{\sigma}^{j} d\sigma \wedge d\tau \notag
\end{align}
Afterwards these two expressions can be inserted in the boundary
condition (\ref{equation4}) for the $(\tilde{M}, \tilde{g},
\tilde{b})$ to obtain
\begin{align}
- i\tilde{b}_{j}^{i}[(d T_{N})_{k}^{j} + \tilde{\omega}_{l}^{'j}
(T_{N})_{k}^{l} - (T_{N})_{l}^{j}\omega_{k}^{'l}] x^{k} \wedge
d\sigma + i\tilde{b}_{j}^{i} (T_{N})_{l}^{j} x_{\tau}^{l} d\sigma
\wedge d\tau = \notag \\ [(d T_{N})_{k}^{i} +
\tilde{\omega}_{j}^{'i} (T_{N})_{k}^{j} -
(T_{N})_{j}^{i}\omega_{k}^{'j}] x^{k} \wedge d\tau +
i(T_{N})_{j}^{i} b_{k}^{j} x_{\tau}^{k} d\sigma \wedge d\tau
\label{equation26}
\end{align}
In order to better understand the resultant expression we define the
following tensors
\begin{align}
U_{k\tau}^{i} d\tau &= i\tilde{b}_{j}^{i}[(d T_{N})_{k}^{j} +
\tilde{\omega}_{l}^{'j} (T_{N})_{k}^{l} -
(T_{N})_{l}^{j}\omega_{k}^{'l}] x^{k} + i \frac{1}{2}
\tilde{b}_{j}^{i} (T_{N})_{k}^{j} x_{\tau}^{k} d\tau - i \frac{1}{2}
(T_{N})_{j}^{i} b_{k}^{j} x_{\tau}^{k} d\tau
\label{equation27}\\
U_{k\sigma}^{i} d\sigma &= [(d T_{N})_{k}^{i} +
\tilde{\omega}_{j}^{'i} (T_{N})_{k}^{j} -
(T_{N})_{j}^{i}\omega_{k}^{'j}) x^{k} - i \frac{1}{2}
\tilde{b}_{j}^{i} (T_{N})_{k}^{j} x_{\tau}^{k} d\sigma + i
\frac{1}{2} (T_{N})_{j}^{i} b_{k}^{j} x_{\tau}^{k} d\sigma
\label{equation28}
\end{align}
These tensors can be put in (\ref{equation26}) to yield
$U_{k\tau}^{i} = U_{k\sigma}^{i} $. Note that these are the constant
tensors. We pull off the minimal case and take them to be zero. It
is manifest that splitting $d T_{N} + \tilde{\omega}^{'} T_{N} -
T_{N}\omega^{'} =\mathcal{ A}^{'} d\sigma + \mathcal{B}^{'} d\tau$
into $d\sigma$ and $d\tau$ directions, and adding and subtracting
(\ref{equation27}) and (\ref{equation28}) one obtains
\begin{align}
(d T_{N})_{k}^{i} + \tilde{\omega}_{j}^{'i}
(T_{N})_{k}^{j} - (T_{N})_{j}^{i}\omega_{k}^{'j} &= 0 \label{equation29}\\
(T_{N})_{j}^{i} b_{k}^{j} &= \tilde{b}_{j}^{i} (T_{N})_{k}^{j}
\label{equation30}
\end{align}
The first result (\ref{equation29}) may be proceeded by taking the
exterior derivative and considering the integrability conditions to
yield the result that curvatures at the boundaries are the same,
$R_{jkl}^{'i} = \tilde{R}_{jkl}^{'i}$. One may proceed in a similar
way to get an infinite number of relations between covariant
derivatives of curvatures. This is not a surprising result since it
is an extension of the bulk space results. But interesting result
appears in (\ref{equation30}) because it describes that
pseudoduality at boundaries requires the equality of two form fields
$b$ and $\tilde{b}$ while pseudoduality in bulk space demands the
equality of torsions (\ref{equation21}). Because antisymmetric $b$
and $\tilde{b}$-fields are locally defined as opposed to the $H$ and
$\tilde{H}$-fields which are globally defined it is understood that
pseudoduality at boundaries impose the locality constraint while it
is globally defined in the bulk space. This is a natural consequence
of the Stokes' theorem which is used to derive the boundary
pseudoduality expressions.

One may verify these results in case that (\ref{equation6}) and
(\ref{equation7}) are extended to the boundaries with the
restriction (\ref{equation4}). Substitute (\ref{equation6}) and
(\ref{equation7}) in $\tilde{x}_{\sigma}^{i} - i \tilde{b}_{j}^{i}
\tilde{x}_{\tau}^{j} = 0$ on $\tilde{M}$ and use the same boundary
condition $x_{\sigma}^{i} - i b_{j}^{i} x_{\tau}^{j} = 0$ on $M$ to
obtain the result (\ref{equation30}).

Now we consider the pseudoduality in case that there exist mixed
boundary conditions. We introduce the projection operators
$\mathcal{P}_{\pm j}^{i} \equiv \frac{1}{2} (\delta_{j}^{i} \pm
\mathcal{R}_{j}^{i})$ as in (\cite{koerber}) where the $(1,
1)$-tensor $\mathcal{R}_{j}^{i} (x)$ satisfies
\begin{equation}
\mathcal{R}_{k}^{i} \mathcal{R}_{j}^{k} = \delta_{j}^{i} \notag
\end{equation}
and leads the metric to be invariant
\begin{equation}
\mathcal{R}_{i}^{j} g_{jk} \mathcal{R}_{l}^{k} = g_{il} \notag
\end{equation}
after all it is a symmetric tensor, $\mathcal{R}_{ij} =
\mathcal{R}_{ji}$. In other words $\mathcal{P}_{+}$ and
$\mathcal{P}_{-}$ project onto the Neumann and Dirichlet directions
respectively. Therefore the boundary conditions (\ref{equation3})
and (\ref{equation4}) can be interpreted as
\begin{align}
&\mathcal{P}_{- j}^{i} \delta x^{j} = 0 \notag\\
&\mathcal{P}_{+ j}^{i} (x_{\sigma}^{j} - i b_{k}^{j} x_{\tau}^{k}) =
0 \notag
\end{align}
These equations can also be expressed as follows
\begin{align}
&\delta x^{i} = \mathcal{P}_{+ j}^{i} \delta x^{j}
\label{equation31}\\ &x_{\sigma}^{i} = \mathcal{P}_{- j}^{i}
x_{\sigma}^{j} + i \mathcal{P}_{+ j}^{i} b_{k}^{j} \mathcal{P}_{+
l}^{k} x_{\tau}^{l} \label{equation32}
\end{align}
where $x_{\tau}^{i} = \mathcal{P}_{+ j}^{i} x_{\tau}^{j}$ is
implemented in (\ref{equation32}) if (\ref{equation31}) can be put
into $\delta x^{i} = x_{\tau}^{i} \delta \tau$ if the time
independence is assumed. Furthermore, it is easy to obtain that the
projection operator $\mathcal{P}_{+}$ is integrable,
\begin{equation}
\mathcal{P}_{+ [i}^{j} \mathcal{P}_{+ k ]}^{l} \mathcal{P}_{+ j,
l}^{m} = 0 \notag
\end{equation}
This integrability condition requires that the commutator of two
infinitesimal displacement in the Neumann direction remains in the
Neumann direction, see (\cite{koerber}). Therefore setting $T = T_M$
and introducing $\mathcal{S}_{\mp \mp \ell}^{i} := \mathcal{P}_{\mp
j}^{i} (T_M)_{k}^{j} \mathcal{P}_{\mp \ell}^{k}$ the boundary
pseudoduality expression (\ref{equation24}) can be written in
Dirichlet and Neumann directions respectively
\begin{align}
\mathcal{\tilde{P}}_{+ j}^{i} \tilde{x}^{j} = \mathcal{S}_{+ +
\ell}^{i} x^{l} \label{equation33}\\
\mathcal{\tilde{P}}_{- j}^{i} \tilde{x}^{j} = \mathcal{S}_{- -
\ell}^{i} x^{l} \label{equation34}
\end{align}
with the requirements
\begin{equation}
\mathcal{S}_{+ - \ell}^{i} x^{l} = 0 \ \ \ \ \ \ \ \ \ \
\texttt{and} \ \ \ \ \ \ \ \ \ \ \mathcal{S}_{- + \ell}^{i} x^{l} =
0 \notag
\end{equation}
Notice that the first expression (\ref{equation33}) represents the
Dirichlet whereas the second one (\ref{equation34}) corresponds to
the Neumann boundary conditions. As a result taking $\delta$ of the
first expression leads to two distinct relations
\begin{align}
\mathcal{\tilde{P}}_{+ j}^{i} \delta \tilde{x}^{j} = \mathcal{S}_{++j}^{i} \delta x^{j}  \label{equation35}\\
\mathcal{\tilde{P}}_{+ j,k}^{i} (T_M)_{\ell}^{j}
\mathcal{S}_{++n}^{k} = \mathcal{S}_{++\ell,m}^{i} \mathcal{P}_{+
n}^{m} \label{equation36}
\end{align}
Note that if $T_M$ = constant is picked, then
$\mathcal{S}_{++\ell}^{i}$ turns to $\mathcal{P}_{+\ell}^{i}$ and
these equations are reduced to the result found above
(\ref{equation24}). Consequently, pseudoduality causes the dirichlet
boundaries to shift by a constant parameter with the condition that
the bulk volume remains unchanged \footnote{Notice that
(\ref{equation36}) gives $\mathcal{\tilde{P}}_{+ j, k}^{i} =
\mathcal{P}_{+ j, k}^{i} $.}.

Now consider the Neumann direction and take the $\sigma$-derivative
of (\ref{equation34}) and use (\ref{equation32}) to obtain the
following results
\begin{align}
\mathcal{\tilde{P}}_{- j}^{i} \tilde{x}_{\sigma}^{j} = (T_M)_{j}^{i}
\mathcal{P}_{- k}^{j} x_{\sigma}^{k} + (T_M)_{j,m}^{i}
\mathcal{P}_{- k}^{m} x_{\sigma}^{k} x^{j}\label{equation37}\\
\tilde{\mathfrak{b}}_{\ell}^{i} \tilde{x}_{\tau}^{\ell} =
(T_M)_{j}^{i} \mathfrak{b}_{\ell}^{j} x_{\tau}^{\ell} +
(T_M)_{j,m}^{i}\mathfrak{b}_{\ell}^{m} x_{\tau}^{\ell} x^{j}
\label{equation38}
\end{align}
where $\mathfrak{b}_{\ell}^{i} := \mathcal{P}_{+ j}^{i} b_{k}^{j}
\mathcal{P}_{+ \ell}^{k}$ is defined. From the first result
(\ref{equation37}) one obtains
\begin{align}
\mathcal{\tilde{P}}_{- j}^{i} (T_M)_{k}^{j} = (T_M)_{j}^{i}
\mathcal{P}_{-
k}^{j} \label{equation39}\\
\mathcal{\tilde{P}}_{- j}^{i} (T_M)_{\ell,k}^{j} =
(T_M)_{\ell,j}^{i} \mathcal{P}_{- k}^{j} \label{equation40}
\end{align}
and the second result (\ref{equation38}) yields
\begin{align}
\tilde{\mathfrak{b}}_{j}^{i} (T_M)_{k}^{j} = (T_M)_{j}^{i} \mathfrak{b}_{k}^{j} \label{equation41}\\
\tilde{\mathfrak{b}}_{j}^{i} (T_M)_{k, \ell}^{j} = (T_M)_{k, j}^{i}
\mathfrak{b}_{\ell}^{j} \label{equation42}
\end{align}
These are all the relations that determine the boundary
pseudoduality equations in case of mixed boundary conditions. For
the trivial case where $T_M$ is a constant, pseudoduality equations
in Dirichlet and Neumann directions simply become
$\mathcal{\tilde{P}}_{\pm j}^{i} \tilde{x}^{j} = \mathcal{P}_{\pm
j}^{i} x^{j}$ with the following conditions
\begin{align}
\mathcal{\tilde{P}}_{+ j}^{i} \delta \tilde{x}^{j} &= \mathcal{P}_{+
j}^{i} \delta x^{j} \ \ \ \ \ \ \ \ \ \
\mathcal{\tilde{P}}_{+ j, k}^{i} = \mathcal{P}_{+ j, k}^{i} \notag\\
\mathcal{\tilde{P}}_{- j}^{i} &= \mathcal{P}_{- j}^{i} \ \ \ \ \ \ \
\ \ \ \ \tilde{\mathfrak{b}}_{j}^{i} = \mathfrak{b}_{j}^{i} \notag
\end{align}
Notice that Dirichlet projection operator is preserved while Neumann
projection operator satisfies the conditions in the first line.
Locality constraint is obvious and given by the equality of
``projected'' antisymmetric $\mathfrak{b}$-fields.

\section{Pseudoduality in WZW Models}
Analysis we established in above sections can be carried out for
sigma models based on group manifolds. We emphasize that group
manifolds are mappings from Euclidean worldsheets. This fixes the
problems we encountered in supersymmetric cases in
\cite{msarisaman1}. Let us consider a strict WZW sigma model
\cite{witten} based on a compact Lie group $G$ of dimension $n$.
Lagrangian of this model is given by
   \begin{equation}
   \mathcal{L} = \frac{1}{2} Tr (\partial_{\mu} g^{-1} \partial^{\mu}
   g) + \Gamma \label{equation43}
   \end{equation}
where $\Gamma$ represents the WZ term, and the field $g(\tau,
\sigma)$ defined on the Euclidean space (possibly with $-\infty <
\sigma \leq 0$)\cite{mackay} takes values in a compact classical Lie
Group $G$ and is given by the map $g : \Sigma \rightarrow G$. There
is a global continuous $G \times G$ symmetry $g \rightarrow U g
V^{-1}$, $U, V \in G$, which gives
$\mathfrak{g}$-valued\footnote{$\mathfrak{g}$ is the Lie algebra of
$G$ with a negative-definite invariant inner product $<\cdot ,
\cdot>$.} conserved currents with zero curvature
    \begin{equation}
    j_{\mu}^{R} = g^{-1} \partial_{\mu} g , ~~~~~~~~j_{\mu}^{L} = - \partial_{\mu}
    g g^{-1} \label{equation44}
    \end{equation}
The equations of motion in bulk space are $\partial^{\mu}
j_{\mu}^{R} = \partial^{\mu} j_{\mu}^{L} = 0$. The boundary equation
of motion at $\sigma = 0$ is
    \begin{equation}
    Tr(g^{-1} \partial_{\sigma} g g^{-1} \delta g) = 0
    \label{equation45}
    \end{equation}
where $g^{-1} \delta g \in \mathfrak{g}$. Obviously one obtains the
Dirichlet and Neumann boundary conditions respectively
    \begin{align}
    \delta g = 0, ~~~~~~~~\partial_{\sigma} g = 0 \label{equation46}
    \end{align}
Notice that boundary equation of motion is equivalent to
$Tr(j_{\tau}, j_{\sigma}) = 0$. We know that both currents generate
the orthonormal coframes $\{j\}$ on the pullback bundle $g^{*}(TG)$
such that they satisfy the Maurer-Cartan equation
    \begin{align}
    dj^{i} + \frac{1}{2} f_{k \ell}^{i} j^{k} \wedge j^{\ell} = 0 \notag
    \end{align}
where $\theta^{i} = j^{i}$, $j^{i} = j_{a}^{i} d\pi^{a} = (g^{-}
\partial_{a} g)^{i} d\pi^{a}$ and $\omega_{\ell}^{i} = \frac{1}{2} f_{k \ell}^{i}
j^{k}$ is the antisymmetric riemannian connection. These coframes
together with the corresponding riemannian connection satisfy the
Cartan structural equations (\ref{equation9})-(\ref{equation10}).
The bulk space pseudoduality equations in the first case are given
by $\tilde{j}^{R}= T j^{R}$ and expressed in the following forms
    \begin{align}
    \tilde{j}_{\tau}^{R} = T j_{\tau}^{R} \label{equation47}\\
    \tilde{j}_{\sigma}^{R} = T j_{\sigma}^{R} \label{equation48}
    \end{align}
We inquire the integrability conditions by taking $\partial_{\tau}$
derivative of (\ref{equation47}), $\partial_{\sigma}$ derivative of
(\ref{equation48}) and adding them together to obtain
    \begin{equation}
     T^{-1}(\partial_{\mu} T) j_{\mu}^{R} = T^{-1}(\partial_{\tau} T) j_{\tau}^{R} + T^{-1}(\partial_{\sigma} T)
     j_{\sigma}^{R} = 0 \label{equation49}
    \end{equation}
Obviously $T$ depends on currents $j_{\tau}^{R}$ and
$j_{\sigma}^{R}$ nonlinearly and solution requires using the
identity \cite{msarisaman4}
     \begin{equation}
     T^{-1}(\partial_{\mu} T) = \frac{1 - e^{-ad X}}{ad X}
     \partial_{\mu} X = \sum_{k=0}^{\infty} \frac{(-1)^{k}}{(k +
     1)!}[X,...[X, \partial_{\mu}X]] \notag
     \end{equation}
where we introduced an exponential solution $T = e^{X}$ \footnote{$X
\in so(n)$, the Lie algebra of $SO(n)$.}, and $ad X : \mathfrak{g}
\rightarrow \mathfrak{g}$, the adjoint representation of $X$, and
$ad X(Y ) = [X, Y]$, $\forall Y \in \mathfrak{g}$. We let $X
\rightarrow \epsilon X$ for small parameter $\epsilon$ and look for
a perturbation solution to get
      \begin{equation}
     (\partial_{\tau} X) j_{\tau}^{R} + (\partial_{\sigma} X)
     j_{\sigma}^{R} = 0 \notag
      \end{equation}
in the first order of $\epsilon$. Trivial solution is that $X$ is a
constant so that T may be chosen to be identity. Therefore,
pseudoduality maps the group manifold $G$ to itself. The general
solution requires tedious analysis, and we just consider a
restricted solution for simplicity. Assume that both terms of
partial differential equation is independent of each other so that
solution for the lie algebra valued $X$ can be written as
     \begin{equation}
     X = \{\int j_{\tau}^{-1} d\tau \cup \int j_{\sigma}^{-1} d\sigma
     \} \label{equation50}
     \end{equation}
where we dropped the upper label $R$ for convenience. Therefore,
infinite number of pseudodual currents can be written in terms of
nonlocal currents using the pseudoduality expressions
(\ref{equation47}) and (\ref{equation48})
     \begin{equation}
     \tilde{j}_{\mu} = \sum_{n=1}^{\infty} \epsilon^{n}
     \tilde{j}_{\mu}^{(n)} \label{equation51}
     \end{equation}
Another intriguing result of special importance is the commutation
relations between currents living on pseudodual manifold and is
obtained by taking $\partial_{\sigma}$ of
(\ref{equation47}),$\partial_{\tau}$ of (\ref{equation48}) and
subtracting from each other
     \begin{equation}
     [\tilde{j}_{\tau}, \tilde{j}_{\sigma}]_{\tilde{G}} = T [j_{\tau}, j_{\sigma}]_{G} +
     (\partial_{\sigma} T) j_{\tau} - (\partial_{\tau} T) j_{\sigma}
     \label{equation52}
     \end{equation}
where $[\cdot , \cdot ]_{G}$ and $[\cdot , \cdot ]_{\tilde{G}}$ are
bracket relations in $G$ and $\tilde{G}$ respectively. Once we find
the solution for $T$ using (\ref{equation50}), we insert in
(\ref{equation52}) and come up with the bracket relation on the
pseudodual manifold $\tilde{G}$.

In the second case pseudoduality equations are expressed by
$\tilde{j} = _{\ast \Sigma} Tj$ and are written explicitly as
       \begin{align}
       \tilde{j}_{\tau} = T j_{\sigma} \label{equation69}\\
    \tilde{j}_{\sigma} = -T j_{\tau} \label{equation70}
       \end{align}
where $j$ stands for both $j^{R}$ and $j^{L}$. Taking
$\partial_{\tau}$ of (\ref{equation69}), $\partial_{\sigma}$ of
(\ref{equation70}) and adding together yields the following
commutation relation which gives rise to a solution for $T$
       \begin{equation}
       [j_{\sigma}, j_{\tau}]_{G} = (T^{-1} \partial_{\sigma}
       T)j_{\tau} - (T^{-1} \partial_{\tau}
       T)j_{\sigma} \label{equation71}
       \end{equation}
This is just a special case of (\ref{equation52}) when the
commutation relation of the pseudodual currents vanish, i.e.
$[\tilde{j}_{\tau}, \tilde{j}_{\sigma}]_{\tilde{G}} = 0$. Likewise
one takes $\partial_{\sigma}$ of (\ref{equation69}),
$\partial_{\tau}$ of (\ref{equation70}) and subtract to get a
commutation relation of pseudodual currents
       \begin{equation}
       [\tilde{j}_{\tau}, \tilde{j}_{\sigma}]_{\tilde{G}} = (\partial^{\mu}T)j_{\mu} = (\partial_{\sigma}
       T)j_{\sigma} + (\partial_{\tau}
       T)j_{\tau} \label{equation72}
       \end{equation}
Hence, using above expansion for $T^{-1}\partial_{\mu}T$, one finds
out a solution of $T$ in (\ref{equation71}) in terms of currents
$j_{\tau}$, $j_{\sigma}$ and commutation relation $[j_{\sigma},
j_{\tau}]_{G}$ and puts this solution in (\ref{equation72}) to
obtain a commutation relation of the pseudodual currents as an
infinite number of series in terms of currents on the generic
manifold $M$. In fact, case II pseudoduality generates an infinite
number of commutation relations while case I pseudoduality just
yields an infinite number of currents in terms of currents on
manifold $M$. To find certain expressions for commutation relations,
a specific solution should be chosen.

Pseudoduality can be extended to boundaries using Stokes' theorem as
above section to obtain the boundary pseudoduality expression in a
simple form
     \begin{equation}
     \tilde{g} = T g \label{equation53}
     \end{equation}
It is obvious that if there is only Dirichlet boundary condition,
then T is trivial, and identity. If we only have Neumann boundary
condition, then taking $\partial_{\sigma}$ yields that $T$ only
depends on $\tau$. It can be any $\tau$-dependent function so that
pseudoduality conditions are satisfied. Therefore, one obtains the
currents at boundaries
     \begin{equation}
     \tilde{j}_{\sigma} = j_{\sigma}, ~~~~~~~~\tilde{j}_{\tau} =
     j_{\tau} + g^{-1} (T^{-1} \partial_{\tau} T) g ~~~~~~\textrm{at~boundary}~(\textrm{at}~\sigma = 0) \notag
     \end{equation}
In the presence of mixed boundary conditions one needs to perform
analysis on the symmetric spaces. This can be accomplished using
above reasoning and results in \cite{msarisaman4}.

\section{Concluding Remaks}

We performed all possible pseudoduality transformations between
different sigma models with Euclidean signatures and extended our
analysis to boundaries. We have seen that there could be two
different types of pseudodualities, each of which produced
intriguing results reflecting their own peculiarities. In case of
pseudoduality producing coordinate diffeomorphism, Integrability
conditions led us to define modified connections $\xi$ and
$\tilde{\xi}$ on manifolds $M$ and $\tilde{M}$ respectively. These
connections provided to find out a general pseudoduality condition
(\ref{equa23}), which yielded a relation between curvatures with
respect to the modified connections. We also obtained infinitely
many torsion relations and their covariant derivatives as given in
(\ref{equation21}), (\ref{equa26}) and (\ref{equa27}). In the
special case that these modified connections vanish, we found out
that torsions and curvature tensors are preserved under
pseudoduality and produce the coordinate diffeomorphisms. Case II
pseudoduality more likely concerns the geometry of the manifolds and
resulted in a conclusion about the geometry of manifolds $M$ and
$\tilde{M}$ to be dual symmetric spaces with respect to modified
connections $\xi$ and $\tilde{\xi}$ respectively. This results in a
conclusion that pseudoduality imposes the manifolds be symmetric
spaces with opposite curvatures. This is the generic feature of case
II pseudoduality, not just special to sigma models due to
(\ref{equation35}) and (\ref{equation37}). This type of
pseudoduality does not allow manifolds $M$ and $\tilde{M}$
torsionful.

We have demonstrated that boundary pseudoduality gives the locality
constraint and preserves the antisymmetric two-form field. Boundary
pseudoduality analysis leads to a convenient framework for the
pseudoduality of D-Branes.

Sigma models based on group manifolds yield an infinite number of
nonlocal conserved currents under case I pseudoduality
(\ref{equation51}). We also obtained the commutation relations
between manifolds $G$ and $\tilde{G}$ (\ref{equation52}). Case II
pseudoduality leads to appropriate commutation relations of currents
on both manifolds $M$ and $\tilde{M}$. Commutation relations on
$\tilde{M}$ are expressed by infinite number of terms as functions
of currents and their commutation relations on manifold $M$.
Boundary conditions are used to find currents at boundaries.

In general, since pseudoduality is performed on the worldsheets
integrability conditions are determined by the metric of worldsheet.
It turns out that Euclidean metric yields well-defined results
compared to Lorentzian metric when worldsheet is supersymmetrized.
It is also intriguing to construct pseudoduality on worldsheets with
a general metric. We plan to explore this case in a more general
context.
\appendix
\section{Appendix}
Lorentzian action corresponding to (\ref{equation1}) is given by
   \begin{equation}
S = \int d\tau d\sigma (\frac{1}{2} g_{ij} \frac{\partial x^{i}}
{\partial \tau} \frac{\partial x^{j}} {\partial \tau} - \frac{1}{2}
g_{ij} \frac{\partial x^{i}} {\partial \sigma} \frac{\partial x^{i}}
{\partial \sigma}) + b_{ij} \frac{\partial x^{i}} {\partial \tau}
\frac{\partial x^{j}} {\partial \sigma}) \label{equationa1}
\end{equation}
where the functions $x^{i}(\sigma)$ giving the dependence of the
real coordinates $x^{i}$ of $M$ on the real coordinates
$\sigma^{\mu}$ of $\Sigma$. The worldsheet $\Sigma$ is endowed with
the Lorentzian metric $h_{\mu \nu}$. Notice that  this Lorentzian
action is real. The bulks space equations of motion following from
this action will be
   \begin{equation}
x_{\sigma \sigma}^{i} - x_{\tau \tau}^{i} = - \Gamma_{jk}^{i}
(x_{\tau}^{j} x_{\tau}^{k} - x_{\sigma}^{j} x_{\sigma}^{k}) -
H_{jk}^{i} x_{\sigma}^{j} x_{\tau}^{k} \label{equationa2}
   \end{equation}
with the corresponding Dirichlet and Neumann boundary conditions
respectively
\begin{align}
\delta x^{i} = 0 \label{equationa3}\\
x_{\sigma}^{i} - b_{j}^{i} x_{\tau}^{j} = 0 \label{equationa4}
\end{align}
Pseudoduality equations are stated with $\tilde{\theta} = _{\ast
\Sigma}T\theta$, where $_{\ast \Sigma}$ denotes the Hodge duality
operator, and given by the following pairs of equations
   \begin{align}
   \tilde{x}_{\tau} = T x_{\sigma} \notag\\
   \tilde{x}_{\sigma} = T x_{\tau} \notag
   \end{align}
Therefore, "particle-like" solutions ($\sigma$-independent) on $M$
get mapped into static "soliton-like" solutions on $\tilde{M}$ and
vice-versa. Integrability conditions for these equations yield that
torsions of both manifolds $M$ and $\tilde{M}$ vanish, and when
extended to supersymmetry, pseudoduality transformation is not
invertible and not well-defined globally \cite{msarisaman1,
msarisaman2, msarisaman3, msarisaman5}.

\bibliographystyle{amsplain}

\end{document}